# Activation cross-sections of longer lived radioisotopes of proton induced nuclear reactions on terbium up to 65 MeV

F.Tárkányi[1], A. Hermanne[2], F. Ditrói[1*], S. Takács[1], A. V. Ignatyuk[3]

[1] *Institute for Nuclear Research, Hungarian Academy of Sciences (ATOMKI), Debrecen, Hungary*

[2] *Cyclotron Laboratory, Vrije Universiteit Brussel (VUB), Brussels, Belgium*

[3] *Institute of Physics and Power Engineering (IPPE), Obninsk, Russia*

## Abstract

Experimental cross sections are presented for the $^{159}$Tb(p,xn)$^{153,155,157,159}$Dy, $^{152,153,155,156m2,m1,g,158}$Tb and $^{153,151}$Gd nuclear reactions up to 65 MeV. The experimental results are compared with the recently reported experimental data and with the results of the nuclear reaction codes ALICE-IPPE, EMPIRE and TALYS as reported in the TENDL-2015 on-line library. Integral thick-target yields are also derived for the reaction products used in practical applications and production routes are discussed.

Keywords: terbium; proton irradiation; theoretical model codes; cross section; physical yield

[*] Corresponding author:ditroi@atomki.hu

# Introduction

A systematic study of proton and deuteron induced reactions up 50 MeV deuteron and 65-70 MeV proton is in progress for testing the prediction possibilities of different theoretical model codes, to complete the activation data file and to investigate production routes of medically relevant isotopes (diagnostic and therapeutic nuclear medicine) of the lanthanide group. Nearly all elements from lanthanum to lutetium were already investigated by our group and discussed in earlier publications.

Activation cross section on Tb are required for production of the medically relevant $^{159}$Dy (Nayak et al., 1999) and $^{157}$Dy (Lebowitz et al., 1971, Yanao et al. 1972, Apo, 1981. Our results on deuteron induced reactions on terbium were published by Tarkanyi et al., 2013. We decided to continue the investigations to compare the production possibilities of the proton and deuteron induced routes.

When our experiment started only a few experimental activation cross-sections of proton induced reaction on terbium were available in the literature, by Lebowitz et al.,1971, Hassan et al., 2010 and Steyn et al. ,2014. Very recently a detailed set of experimental data up to 200 MeV was published by the Los Alamos group (Engle et al. 2012,2012b,2015,2016).

# Experimental and data evaluation

The experimental details and the used data evaluation methods were described in many of our earlier reports (Ditrói et al, 1991, Ditrói et al, 2014). Here we briefly summarize the parameters closely related to this study in Table 1 (experiments), Table 2 (data evaluation) and Table 3 (used decay data), respectively.

Table 1. Main experimental parameters

| Reaction | $^{159}$Tb(p,x) (ser. 2) | $^{159}$Tb(p,x) (ser. 1) |
|---|---|---|
| Incident particle | Proton | Proton |
| Method | Stacked foil | Stacked foil |
| Number of Tb target foils | 15 | 12 |
| Target composition and thickness (μm) | Tb(22.2), Al(377), La(25), Al(10), CeO(33.1-63.2, sedimented), Al (100) Repeated 15 times | Ti (10.9), Al (102), Tb (22.2). Ti (10.9), Al (102), Al (10) CeO(4.59-34.0, sedimented), Al (100) Repeated 12 times |
| Accelerator | Cyclone 90 cyclotron of the Université Catholique in Louvain la Neuve (LLN) | CGR-560 cyclotron of the Vrije Universiteit Brussel (VUB) in Brussels |
| Primary energy (MeV) | 65 | 35 |
| Covered energy range (MeV) | 64.9-35.3 | 33.9-5.8 |
| Irradiation time (min) | 59 | 60 |
| Beam current (nA) | 90 | 100 |
| Monitor reaction, [recommended values] | $^{27}$Al(p,x)$^{24}$Na reaction (Tarkanyi et al., 2001) | $^{nat}$Ti(p,x)$^{48}$V reaction (Tarkanyi et al., 2001) |
| Monitor target and thickness (μm) | $^{nat}$Al, 377 and (100+10) | $^{nat}$Ti, 10.9 |
| detector | HPGe | HPGe |
| γ-spectra measurements | 4 series | 4 series |
| Cooling times (h) | 6.4-8.9 | 1.4-3.8 |
| | 22.2-28.1 | 22.1-25.1 |
| | 127.0-166.1 | 292-319 |
| | 3120- 3241 | 1998-2283 |

Table 2. Main parameters of data evaluation (with references)

| Gamma spectra evaluation | Genie 2000, Forgamma | (Genie 2000, Szekely, 1985) |
|---|---|---|
| Determination of beam intensity | Faraday cup (preliminary) | |
| | Fitted monitor reaction (final) | (Tarkanyi et al.,1991) |
| Decay data | NUDAT 2.6 | (Kinsey et al 1997) |
| Reaction Q-values | Q-value calculator | (Pritychenko and Sonzogni) |
| Determination of beam energy | Anderson (preliminary) | (Anderson and Ziegler, 1977) |
| | Fitted monitor reaction (final) | (Tarkanyi et al., 1991) |
| Uncertainty of energy | Cumulative effects of possible uncertainties | |
| Cross sections | Isotopic cross section | |
| Uncertainty of cross sections | Sum in quadrature of all individual linear contributions | (Guide) |
| Yield | Physical yield | (Bonardi ,1988, Otuka and Takacs, 2015) |

Table. 3 Investigated radionuclides, their decay properties and production routes (Kinsey et al. 1997, Pritychenko et al.)

| Nuclide | Half-life | $E_\gamma$(keV) | $I_\gamma$ (%) | Contributing reaction | Q-value (keV) |
|---|---|---|---|---|---|
| **$^{159}$Dy** <br> ε: 100 % | 144.4 d | 58.0 | 2.27 | $^{159}$Tb(p,n) | -1147.72 |
| **$^{157}$Dy** <br> ε: 100 % | 8.14 h | 182.424 <br> 326.336 | 1.33 <br> 93 | $^{159}$Tb(p,3n) | -17033.0 |
| **$^{155}$Dy** <br> ε: 100 % | 9.9 h | 184.564 <br> 226.918 | 3.37 % <br> 68.4 | $^{159}$Tb(p,5n) | -33444.3 |
| **$^{153}$Dy** <br> α: 0.0094 % <br> ε: 99.9906 % | 6.4 h | 80.723 <br> 99.659 <br> 213.754 <br> 254.259 <br> 274.673 <br> 389.531 <br> 537.225 <br> 1023.99 | 11.1 <br> 10.51 <br> 10.9 <br> 8.6 <br> 3.1 <br> 1.52 <br> 1.33 <br> 1.09 | $^{159}$Tb(p,7n) | -49599.33 |
| **$^{156m2}$Tb** <br> IT: 100 % <br> 88.4 keV | 5.3 h | 88.4 | 1.15 | | |
| **$^{156m1}$Tb** <br> IT: 100 % <br> 49.630+X keV | 24.4 h | 49.630 | 74.1 | | |
| **$^{156}$Tb** <br> ε: 100 % | 5.35 d | 88.97 <br> 199.19 | 18 <br> 41 | $^{159}$Tb(p,p3n) | -23655.51 |

| | | 262.54 | 5.8 | | |
| | | 296.49 | 4.5 | | |
| | | 356.38 | 13.6 | | |
| | | 422.34 | 8.0 | | |
| | | 534.29 | 67 | | |
| | | 1065.11 | 10.8 | | |
| | | 1154.07 | 10.4 | | |
| | | 1159.03 | 7.2 | | |
| | | 1222.44 | 31 | | |
| | | 1421.67 | 12.2 | | |
| $^{155}$Tb<br>ε: 100 % | 5.32 d | 86.55<br>105.318<br>148.64<br>161.29<br>163.28<br>180.08<br>340.67<br>367.36 | 32.0<br>25.1<br>2.65<br>2.76<br>4.44<br>7.5<br>1.18<br>1.48 | $^{159}$Tb(p,p4n)<br>$^{155}$Dy decay | -30567.5<br>-33444.3 |
| $^{154m2}$Tb<br>ε: **98.2** %<br>IT **1.8** % | 22.7 h | 123.071<br>225.94<br>247.925<br>346.643<br>1419.81 | 43<br>26.8<br>79<br>69<br>46 | $^{159}$Tb(p,5n) | -39732.6 |
| $^{153}$Tb<br>ε: 100 % | 2.34 d | 102.255<br>109.758<br>170.42<br>212.00 | 6.4<br>6.8<br>6.3<br>31.0 | $^{159}$Tb(p,p6n)<br>$^{153}$Dy decay | -46646.62<br>-49599.33 |
| $^{152}$Tb<br>ε: 100 % | 17.5 h | 271.09<br>344.2785<br>586.27<br>778.9045 | 9.53<br>63.5<br>9.21<br>5.54 | $^{159}$Tb(p,p7n)<br>$^{152}$Dy decay | -55314.6<br>-56695.29 |

| Nuclide | Half-life | Eγ (keV) | Iγ (%) | Production | Q-value (keV) |
|---|---|---|---|---|---|
| $^{153}$Gd<br>ε: 100 % | 240.4 d | 97.43100<br>103.18012 | 29.0<br>21.1 | $^{159}$Tb(p,2p5n)<br>$^{153}$Tb decay | -44295.28<br>-46646.62 |
| $^{151}$Gd<br>ε: 100 % | 123.9 d | 153.60<br>174.70<br>243.29 | 6.2<br>2.96<br>5.6 | $^{159}$Tb(d,2p7n)<br>$^{151}$Tb decay | -59131.89<br>-62479.26 |

Increase the Q-values if compound particles are emitted by: np-d, +2.2 MeV; 2np-t, +8.48 MeV; n2p-$^3$He, +7.72 MeV; 2n2p-α, +28.30 MeV.
Decrease Q-values for isomeric states with level energy of the isomer

## Model calculations

The new experimental data are shown together with earlier reported experimental data and results of nuclear reaction model calculations. The ALICE-IPPE (Dityuk et al., 1998) and EMPIRE (Herman et al., 2007) codes were used to analyze the present experimental results. The parameters for the optical model, level densities and pre-equilibrium contributions were taken as described in (Belgya et al., 2006). The cross sections for isomers in case of ALICE-IPPE code were obtained by using the isomeric ratios calculated with EMPIRE. The new experimental data are also compared to the data in the TENDL-2015 library (Koning et al., 2015), based on both default and adjusted TALYS (1.6) calculations (Koning et al., 2012) .

In a first attempt the EMPIRE calculations were performed using the BNL option (EGSM, Enhanced Generalized Superfluid Model) for level densities, but resulted in large disagreement for excitation functions of gadolinium radio-products. A second series of EMPIRE calculations were performed with the Composite Gilbert-Cameron (GC) description for level densities and gives better agreement for gadolinium isotopes as can be seen on the relevant figures. The differences between the two approaches are rather small for the production of Dy and Tb isotopes and are not represented on the figures. Unfortunately, there are no direct experimental data on the level density parameters for these isotopes.

## Results and discussion

### Excitation functions

The measured cross sections for the production of the investigated radio-products are presented in Tables 4-5 and Figures 1–13 in comparison with earlier literature values and the results of different model calculations.

Table 4 Numerical cross-sections values for $^{153,155,157,159}$Dy and $^{158}$Tb production

| E (MeV) | ΔE (MeV) | $^{159}$Dy σ (mb) | Δσ (mb) | $^{157}$Dy σ (mb) | Δσ (mb) | $^{155}$Dy σ (mb) | Δσ (mb) | $^{153}$Dy σ (mb) | Δσ (mb) | $^{158}$Tb σ (mb) | Δσ (mb) |
|---|---|---|---|---|---|---|---|---|---|---|---|
| 33.93 | 0.20 | 6.55 | 0.84 | 546.83 | 61.39 | | | | | 187.81 | 30.03 |
| 31.66 | 0.28 | 7.21 | 0.88 | 708.20 | 79.50 | | | | | 165.70 | 31.78 |
| 29.25 | 0.36 | 12.29 | 1.54 | 1096.36 | 123.07 | | | | | 135.08 | 27.02 |
| 26.79 | 0.44 | 18.75 | 2.13 | 1184.63 | 132.98 | | | | | 125.85 | 31.31 |
| 24.40 | 0.52 | 13.11 | 1.49 | 1156.37 | 129.81 | | | | | 106.45 | 25.79 |
| 22.41 | 0.59 | 13.23 | 1.51 | 874.44 | 98.16 | | | | | 99.75 | 24.42 |
| 20.05 | 0.67 | 15.90 | 1.82 | 473.83 | 53.19 | | | | | 63.04 | 18.54 |
| 17.49 | 0.76 | 24.18 | 2.72 | 63.29 | 7.11 | | | | | | |
| 14.86 | 0.85 | 45.64 | 5.20 | 0.09 | 0.01 | | | | | | |
| 11.16 | 0.97 | | | | | | | | | | |
| 8.15 | 1.08 | 200.64 | 22.55 | | | | | | | | |
| 5.76 | 1.16 | 129.59 | 14.57 | | | | | | | | |
| | | | | | | | | | | | |
| 64.95 | 0.20 | | | 72.16 | 8.11 | 160.43 | 18.01 | 64.42 | 7.56 | | |
| 62.90 | 0.27 | | | 70.80 | 7.97 | 173.33 | 19.46 | 38.04 | 4.94 | | |
| 60.71 | 0.34 | | | 82.04 | 9.23 | 225.68 | 25.34 | 13.00 | 2.47 | | |
| 58.44 | 0.42 | | | 84.93 | 9.55 | 272.59 | 30.60 | 3.59 | 1.65 | | |
| 56.16 | 0.50 | 7.71 | 1.33 | 91.14 | 10.24 | 340.50 | 38.23 | | | 120.82 | 35.10 |
| 55.10 | 0.54 | | | 93.01 | 10.46 | 364.10 | 40.88 | | | | |
| 54.01 | 0.57 | 8.17 | 2.43 | 98.30 | 11.05 | 392.02 | 44.01 | | | 138.00 | 39.47 |
| 52.50 | 0.62 | | | 98.68 | 11.10 | 419.85 | 47.14 | | | | |
| 51.18 | 0.67 | 10.91 | 1.60 | 105.25 | 11.83 | 452.50 | 50.80 | | | 163.56 | 25.32 |
| 49.94 | 0.71 | | | 113.07 | 12.71 | 466.02 | 52.32 | | | | |
| 48.77 | 0.75 | 10.82 | 1.82 | 110.18 | 12.39 | 410.75 | 46.11 | | | 126.02 | 38.35 |
| 47.44 | 0.80 | | | 115.87 | 13.02 | 413.09 | 46.38 | | | | |
| 46.13 | 0.84 | | | 123.89 | 13.92 | 384.94 | 43.22 | | | | |
| 45.34 | 0.87 | | | 131.92 | 14.82 | 362.87 | 40.74 | | | | |
| 44.14 | 0.91 | 10.69 | 1.78 | 140.84 | 15.83 | 308.29 | 34.61 | | | 135.64 | 37.40 |
| 43.32 | 0.94 | | | 147.41 | 16.56 | 263.88 | 29.63 | | | | |
| 42.09 | 0.98 | | | 158.20 | 17.78 | 198.40 | 22.27 | | | | |
| 41.23 | 1.01 | | | 159.75 | 17.95 | 145.33 | 16.32 | | | | |
| 39.89 | 1.05 | | | 185.47 | 20.84 | 91.22 | 10.24 | | | | |
| 39.00 | 1.08 | | | 178.73 | 20.08 | 50.74 | 5.70 | | | | |
| 37.65 | 1.13 | | | 217.85 | 24.47 | 18.20 | 2.05 | | | | |
| 36.71 | 1.16 | | | 257.79 | 28.95 | 10.51 | 1.19 | | | | |
| 35.23 | 1.21 | | | 309.46 | 34.75 | 1.91 | 0.25 | | | | |
| 34.24 | 1.24 | | | 343.40 | 38.56 | | | | | | |

Table 5 Numerical cross-sections values for $^{155,156m2,m1,g}$Tb and $^{153}$Gd production

| | | $^{156m2}$Tb | | $^{156m1}$Tb | | $^{156g}$Tb | | $^{155}$Tb | | $^{153}$Gd | |
|---|---|---|---|---|---|---|---|---|---|---|---|
| E (MeV) | σ (mb) | Δσ (mb) | σ (mb) | Δσ (mb) | σ (mb) | Δσ (mb) | σ (mb) | Δσ (mb) | Δσ (mb) | σ (mb) | Δσ (mb) |
| 33.93 | 0.20 | 8.66 | 1.21 | | | 10.50 | 1.26 | 0.471 | 0.080 | 2.37 | 0.27 |
| 31.66 | 0.28 | 3.61 | 0.41 | | | 4.94 | 0.65 | | | 1.75 | 0.20 |
| 29.25 | 0.36 | 1.46 | 0.16 | | | 1.98 | 0.36 | | | 1.19 | 0.13 |
| 26.79 | 0.44 | | | | | | | | | 0.37 | 0.04 |
| 24.40 | 0.52 | | | | | | | | | 0.20 | 0.02 |
| 22.41 | 0.59 | | | | | | | | | 0.11 | 0.01 |
| 20.05 | 0.67 | | | | | | | | | 0.12 | 0.02 |
| 17.49 | 0.76 | | | | | | | | | 0.06 | 0.01 |
| 14.86 | 0.85 | | | | | | | | | 0.09 | 0.01 |
| 11.16 | 0.97 | | | | | | | | | | |
| 8.15 | 1.08 | | | | | | | | | | |
| 5.76 | 1.16 | | | | | | | | | | |
| | | | | | | | | | | | |
| 64.95 | 0.20 | 15.21 | 7.19 | 26.62 | 3.30 | 122.79 | 13.93 | 316.24 | 35.72 | 69.24 | 8.36 |
| 62.90 | 0.27 | 12.33 | 8.13 | 25.28 | 3.22 | 112.95 | 12.99 | 320.79 | 36.33 | 49.00 | 6.61 |
| 60.71 | 0.34 | 13.77 | 8.54 | 21.32 | 2.86 | 128.37 | 14.68 | 393.00 | 44.41 | 41.33 | 5.98 |
| 58.44 | 0.42 | 17.28 | 9.58 | 19.18 | 2.57 | 124.77 | 14.26 | 446.80 | 50.49 | | |
| 56.16 | 0.50 | | | | | 127.47 | 14.48 | 514.57 | 57.96 | 10.96 | 1.23 |
| 55.10 | 0.54 | 19.02 | 6.67 | 31.18 | 8.67 | 123.04 | 14.05 | 531.76 | 59.96 | | |
| 54.01 | 0.57 | 14.96 | 5.06 | 20.11 | 4.54 | 127.12 | 14.52 | 561.53 | 63.33 | 9.65 | 1.09 |
| 52.50 | 0.62 | | | 26.28 | 6.78 | 117.34 | 13.38 | 591.75 | 66.63 | | |
| 51.18 | 0.67 | | | 31.05 | 8.60 | 119.60 | 13.70 | 614.33 | 69.27 | 12.58 | 1.42 |
| 49.94 | 0.71 | | | 22.02 | 4.43 | 120.92 | 13.79 | 613.43 | 69.09 | | |
| 48.77 | 0.75 | | | 26.11 | 8.65 | 109.47 | 12.55 | 565.68 | 63.77 | 17.73 | 4.48 |
| 47.44 | 0.80 | | | 13.48 | 3.59 | 109.02 | 12.46 | 535.32 | 60.37 | | |
| 46.13 | 0.84 | | | 23.05 | 4.75 | 101.69 | 11.66 | 498.12 | 56.21 | | |
| 45.34 | 0.87 | | | 8.08 | 2.52 | 96.92 | 11.14 | 472.64 | 53.35 | | |
| 44.14 | 0.91 | | | | | 95.24 | 10.94 | 385.50 | 45.55 | | |
| 43.32 | 0.94 | | | | | 87.50 | 10.07 | 342.21 | 38.70 | 11.89 | 3.44 |
| 42.09 | 0.98 | | | | | 74.62 | 8.69 | 241.22 | 27.50 | | |
| 41.23 | 1.01 | | | | | 67.00 | 7.68 | 180.59 | 20.48 | | |
| 39.89 | 1.05 | | | | | 54.36 | 6.40 | 116.82 | 13.50 | 12.45 | 2.49 |
| 39.00 | 1.08 | | | | | 41.42 | 4.91 | 63.65 | 7.51 | | |
| 37.65 | 1.13 | | | | | 34.09 | 3.96 | 50.81 | 11.65 | | |
| 36.71 | 1.16 | | | | | 29.67 | 3.47 | 11.10 | 1.95 | | |
| 35.23 | 1.21 | | | | | 18.96 | 2.18 | | | | |
| 34.24 | 1.24 | | | | | 12.74 | 1.44 | 1.43 | 0.27 | | |

Table 6 Numerical cross-sections values for $^{154m2,153,152}$Tb and $^{151}$Gd production

| E (MeV) | ΔE (MeV) | $^{154m2}$Tb σ (mb) | Δσ (mb) | $^{153}$Tb σ (mb) | Δσ (mb) | $^{152}$Tb σ (mb) | Δσ (mb) | $^{151}$Gd σ (mb) | Δσ (mb) |
|---|---|---|---|---|---|---|---|---|---|
| 64.95 | 0.20 | 6.51 | 0.74 | 107.77 | 12.14 | 10.98 | 3.37 | 5.46 | 0.68 |
| 62.90 | 0.27 | 4.92 | 0.56 | 53.05 | 6.07 | 9.34 | 4.11 | | |
| 60.71 | 0.34 | 3.57 | 0.41 | 23.35 | 2.80 | 7.91 | 2.96 | 6.41 | 0.82 |
| 58.44 | 0.42 | 2.25 | 0.27 | 5.96 | 1.07 | 4.18 | 2.10 | | |
| 56.16 | 0.50 | 0.99 | 0.14 | 3.12 | 0.78 | | | 6.34 | 0.77 |
| 55.10 | 0.54 | 1.05 | 0.14 | | | | | | |
| 54.01 | 0.57 | 0.77 | 0.11 | | | | | 6.49 | 0.83 |
| 52.50 | 0.62 | 0.50 | 0.09 | | | | | | |
| 51.18 | 0.67 | 0.58 | 0.10 | | | | | | |
| 49.94 | 0.71 | 0.34 | 0.08 | | | | | | |
| 48.77 | 0.75 | | | | | | | 5.76 | 0.75 |
| 47.44 | 0.80 | | | | | | | | |
| 46.13 | 0.84 | | | | | | | | |
| 45.34 | 0.87 | | | | | | | | |
| 44.14 | 0.91 | | | | | | | | |
| 43.32 | 0.94 | | | | | | | | |
| 42.09 | 0.98 | | | | | | | | |
| 41.23 | 1.01 | | | | | | | | |
| 39.89 | 1.05 | | | | | | | 5.07 | 0.66 |
| 39.00 | 1.08 | | | | | | | | |
| 37.65 | 1.13 | | | | | | | | |
| 36.71 | 1.16 | | | | | | | | |
| 35.23 | 1.21 | | | | | | | | |
| 34.24 | 1.24 | | | | | | | | |

### Production of $^{159}$Dy (T$_{1/2}$=144.4 d)

According to Fig. 1. our new data are in acceptable agreement with the recent results of Engle et al. (2016) and with earlier measurement of Hassan et al., 2007 and Steyn et al., 2014. in the overlapping energy range. The values near the maximum at 9 MeV differ however by a factor of 2. The TENDL-2015 underestimates the maximum around 10 MeV and overestimates systematically the high energy tail. The EMPIRE and the ALICE-IPPE describe rather well the shape of the excitation function with a strong underestimation for the ALICE-IPPE results.

### Production of $^{157}$Dy (T$_{1/2}$=8.14 h)

The present and the earlier experimental data of Lebowitz et al.,1971, Steyn et al., 2014 and Engle et al., 2016 for the $^{159}$Tb(p,3n)$^{157}$Dy reaction are shown in Fig.2. The agreement in the common energy range is excellent. The TENDL-2015 underestimates the maximum. The predictions of the EMPIRE and ALICE-IPPE are acceptable.

### Production of $^{155}$Dy (T$_{1/2}$=9.9 h)

Good agreement was found with the earlier experimental data of Steyn et al., 2014 and Engle et al., 2016. The maximum of the excitation function in the TENDL-2015 prediction is shifted to lower energy (Fig. 3). There are significant disagreements in the maximum value for the EMPIRE and ALICE-IPPE codes.

### Production of $^{153}$Dy (T$_{1/2}$=5.35 d)

Our data are consistent with Steyn et al. ,2014 and the higher energy data of Engle et al., 2016. There is a significant shift to low energy in the effective threshold of the TENDL-2015 excitation function and a significant underestimation of the maximum in case of EMPIRE results (Fig. 4).

## Production of $^{158}$Tb ($T_{1/2}$ = 180 y)

The experimental and theoretical excitation functions are shown in Fig. 5. In this case our data are more scattered but are not contradictory to Engle et al., 2016. There is a factor of 2 disagreement between the predictions of the maximum values of different model codes but all are confirming the general behavior of the experimental excitation function.

## Production of $^{156m2}$Tb ($T_{1/2}$=24.4 h), $^{156m1}$Tb ($T_{1/2}$=5.3 h) and $^{156}$Tb ($T_{1/2}$=5.35 d) (m+)

The radionuclide $^{156}$Tb has a long-lived ground state and two shorter-lived isomeric states decaying completely to the ground state. We tried to deduce cross-sections for direct production of both isomeric states and for cumulative production of the ground state (after the complete decay of isomers) (Figs. 6,7,8). No earlier experimental data for production of isomeric states are available. Our experimental data for isomeric states contain large uncertainties due to the separation of complex gamma lines at 50 and 88 keV and the need for separation of contributions for overlapping γ-lines. Our experimental data for the isomeric states differ significantly from the predictions of the model codes. For cumulative production of the ground state our data are in good agreement with earlier experimental results of Steyn et al., 2014 and Engle et al., 2016. The EMPIRE results show a strange description around 50 MeV while the two other codes describe well the experimental values.

## Production of $^{155}$Tb ($T_{1/2}$=5.32 d) (cum)

The $^{155}$Tb is produced directly by the $^{159}$Tb(p,p4n) reaction with high threshold and through the decay of the $^{155}$Dy (9.9 h) parent radioisotope. Our data represent cumulative production, deduced from spectra after complete decay of the parent. The new data are lower comparing to cumulative data of Steyn et al., 2014 and Engle et al., 2016. The shape of the excitation function of TENDL-2015 differs significantly from the experimental data (Fig. 9). The predictions of ALICE-IPPE are closer to the earlier experimental data, the EMPIRE values correspond better to our results.

### Production of $^{154m2}$Tb  ($T_{1/2}$=22.7 h) (cum)

Out of the tree long-lived states of $^{154}$Tb ($^{154m2}$Tb: 22 h, $^{154m1}$Tb: 5.3 h and $^{154g}$Tb: 21h) we obtained cross-section data for the higher lying isomeric state only.  Both isomeric states decay partly with internal transition to the ground state and with EC to $^{154}$Gd. The investigated $^{154m2}$Tb is produced only directly. All experimental data for production of $^{154m2}$Tb are in good agreement, but the TENDL-2015 and ALICE-IPPE predictions significantly differ both in shape and in magnitude (Fig. 10) while EMPIRE is rather near the experimental values.

### Production of $^{153}$Tb  ($T_{1/2}$=2.34 d) (cum)

The measured cross-sections for production of $^{153}$Tb are cumulative, measured after complete decay of the $^{153}$Dy (6.4 h) parent isotope. The three experimental datasets confirm each other. A large disagreement with the data in the TENDL-2015 library and with the prediction of EMPIRE is seen, while ALICE-IPPE gives a good description (Fig. 11).

### Production of $^{152}$Tb  ($T_{1/2}$= 17.5 h) (cum)

The measured cross sections for production of $^{152}$Tb are cumulative, measured after complete decay of the shorter-lived ($T_{1/2}$ = 4.2 min)  isomeric state   and of the $^{152}$Dy ($T_{1/2}$ = 2.38 h) parent isotope. The agreement with the earlier experimental data is good. The TENDL-2015 differs significantly from the experimental data.

### Production of $^{153}$Gd  ($T_{1/2}$=240.4 d) (cum)

The experimental data for cumulative production of long-lived $^{153}$Gd are shown in Fig. 13. The experimental data for cumulative formation were deduced from spectra measured after decay of the 153 mass parent decay chain ($^{153}$Dy -  6.4 h, $^{153}$Tb - 2.34 d).  There is an acceptable agreement with the experimental data of Steyn et al., 2014 and Engle et al., 2016. The ALICE-IPPE results describe rather well the excitation function. The predictions of TENDL-2015 are energy shifted above 45 MeV. An important influence of the choice of description of level densities is seen for the EMPIRE calculations: while use of EGSM results in a totally discrepant excitation function the second calculation with the GC formulation gives a description similar to ALICE-IPPE but with somewhat overestimated cross-section values.

**Production of $^{151}$Gd ($T_{1/2}$=120 d) (cum)**

We could calculate the activation cross-section of long-lived $^{151}$Gd only for few points. The data are cumulative, measured after decay of parent nuclides in the $^{151}$Dy (17 min)→ $^{151}$Tb (28s, 17.8 h)→ $^{151}$Gd (150 d) decay chain (Fig. 14). There are large disagreements with the results of the used theoretical codes. Here the use of GC description reduces the overestimation predicted by EMPIRE, but the agreement with experimental values is still far off.

### *Integral yields*

The so called physical integral yield (Bonardi, 1981, Otuka et al., 2016) was calculated from experimental excitation functions (spline fitted to experimental data) and shown in Fig. 14.

## Comparison of the production routes of $^{157}$Dy and $^{159}$Dy

Among the investigated reactions the pathways leading to production of $^{157}$Dy and $^{159}$Dy are presently of practical interest in nuclear medicine (Subramanian et al. 1984, dos Santos Agusto et al., 2014).

The light charged particle induced production methods were discussed in detail in our work dealing with measurement of activation cross section of deuteron induced reactions on terbium (Tarkanyi et al., 2013). Here we summarize only briefly the main production routes.

Both $^{157}$Dy and $^{159}$Dy can be produced by neutron induced reactions via (n,$\gamma$) on Dy isotopes (Rao et al. 1977), by photonuclear reactions (Habs et al., 2011), by high energy spallation reaction (Beyer et al., 2003) using electromagnetic separators and by low and medium energy light charged particle induced nuclear reactions.

The main production routes of $^{157}$Dy and $^{159}$Dy via light charged particle beams include proton and deuteron induced reactions on holmium, terbium, dysprosium and $^3$He-and alpha particle induced reactions on gadolinium. The production cross sections of $^{157}$Dy and $^{159}$Dy on holmium and dysprosium were recently measured by us up to 65 MeV proton energy and 50 deuteron energy (Tarkanyi et al., 2016, Tarkanyi et al., 2014, Ditroi et al., 2015). This route requires non-conventional, higher energy machine up to 100 MeV, and will not be discussed in this paper. For medium and low energy accelerators we already made a comparison, only based on theoretical results, in our paper dealing with deuteron induced reactions on terbium (Tarkanyi et al., 2013). The new experimental data, for protons discussed in this publication, do not change the conclusions

*$^{159}$Dy*. For production of $^{159}$Dy at low energy cyclotrons ($E_{p,max} < 20$ MeV) only the $^{159}$Tb(p,n) process can be used while if higher energy is available also the $^{159}$Tb (d,2n) reaction is an option. As it was shown also in our previous work the yield of the (d,2n) reaction is significantly higher compared to (p,n) on the same target in this mass region. The alpha- and $^3$He-particle induced reactions have many drawbacks: lack of available accelerators, lower production yield due to the high stopping power, low available beam intensity, necessity to use of highly enriched target, high price of $^3$He beam.

***157Dy***: At low energy cyclotrons only the low yield $^{154}$Gd($\alpha$,n) reaction (for $^{155}$Gd($\alpha$,2n) the Q-value is already16.36 MeV) is available. At commercial 30-35 MeV cyclotrons the $^{159}$Tb(p,3n) reaction could be used. It has high yields and the activity of the simultaneously produced long-lived $^{159}$Dy is small (low cross sections of the (p,n) reaction in this energy range). The $^{159}$Tb(d,4n) route has high yield, but it requires even higher energy accelerator and the impurity level of the $^{159}$Dy due to the high yield (d,2n) is more significant. The difficulties with the $^3$He and alpha particle reactions are the same as in case of production of $^{159}$Dy.

## Summary and conclusion

We report new experimental cross sections for the reactions $^{159}$Tb(p,xn)$^{153,155,157,159}$Dy, $^{152,153,155,156m2,m1,g,158}$Tb and $^{151,153}$Gd up to 65 MeV. The new data are measured in more detailed energy steps, are in good agreement with the experimental data in the literature in the overlapping energy range and successfully complete the recent experimental data obtained at higher energy. The agreement with the TENDL-2015 (TALYS), EMPIRE and ALICE-IPPE predictions, taking into account the capabilities of the theory is mediocre, even with adapting the description of level densities.

The possible use of the experimental data for $^{159}$Dy and $^{157}$Dy formation is discussed and we conclude that proton induced reactions on terbium could be effectively used for batch production of $^{157}$Dy.

### Acknowledgments

This work was performed in the frame of the HAS–FWO Vlaanderen (Hungary–Belgium) project. The authors acknowledge the support of the research project and of the respective institutions. We thank to Cyclotron Laboratory of the Université Catholique in Louvain la Neuve (LLN) providing the beam time and the crew of the LLN Cyclone 90 cyclotron for performing the irradiations.

**Figures**

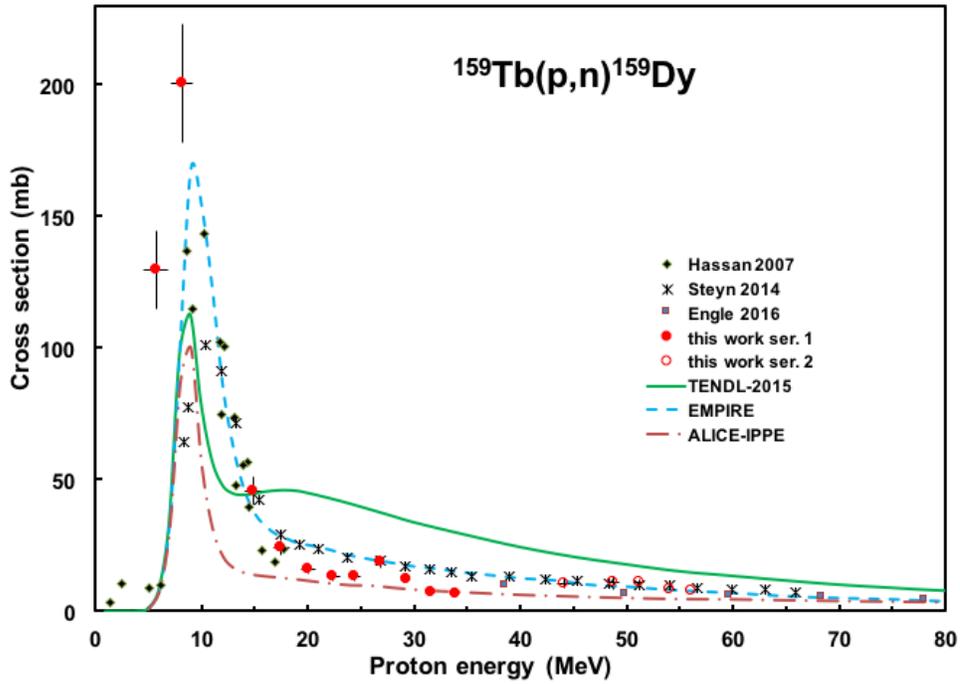

Fig. 1. Experimental and theoretical excitation functions of the $^{159}$Tb(p,n)$^{159}$Dy reaction

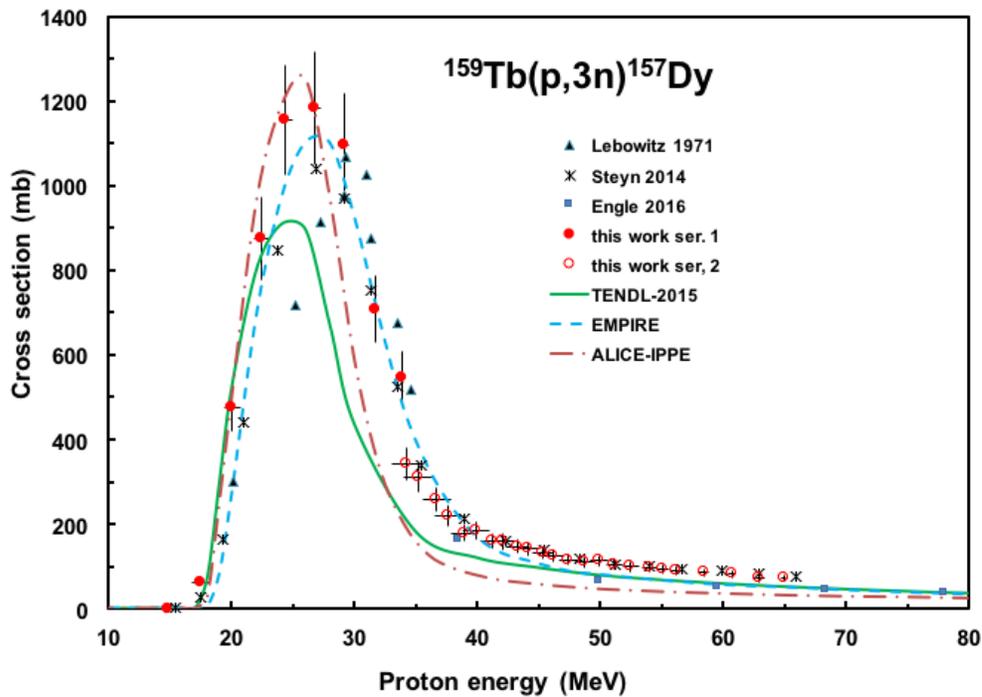

Fig. 2. Experimental and theoretical excitation functions of the $^{159}$Tb(p,3n)$^{157}$Dy reaction

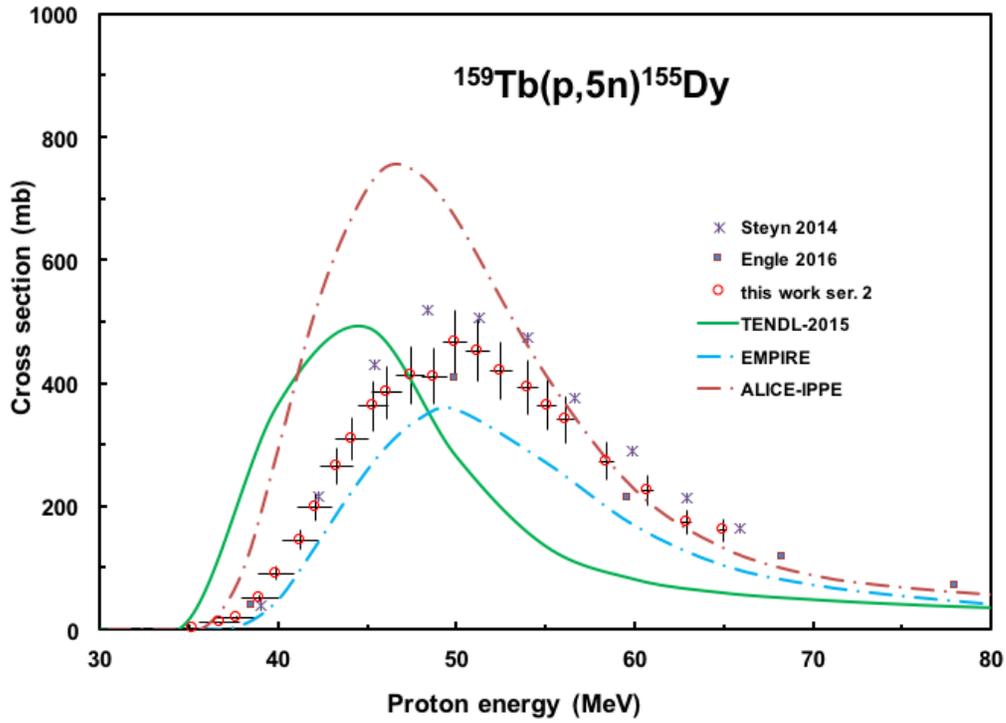

Fig. 3. Experimental and theoretical excitation functions of the $^{159}$Tb(p,5n)$^{155}$Dy reaction

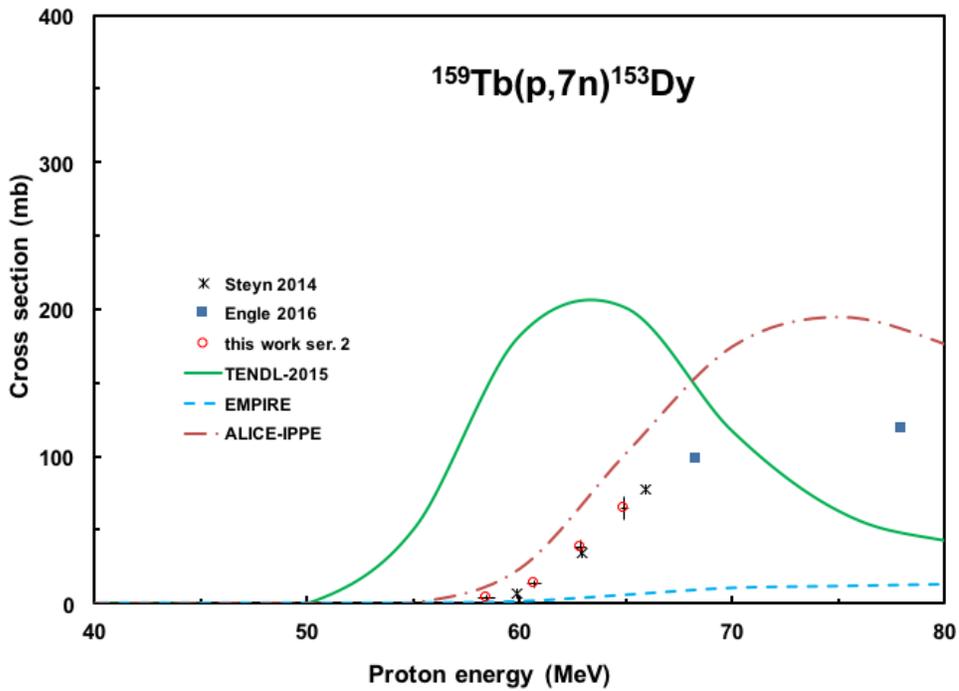

Fig. 4. Experimental and theoretical excitation functions of the $^{159}$Tb(p,7n)$^{153}$Dy reaction

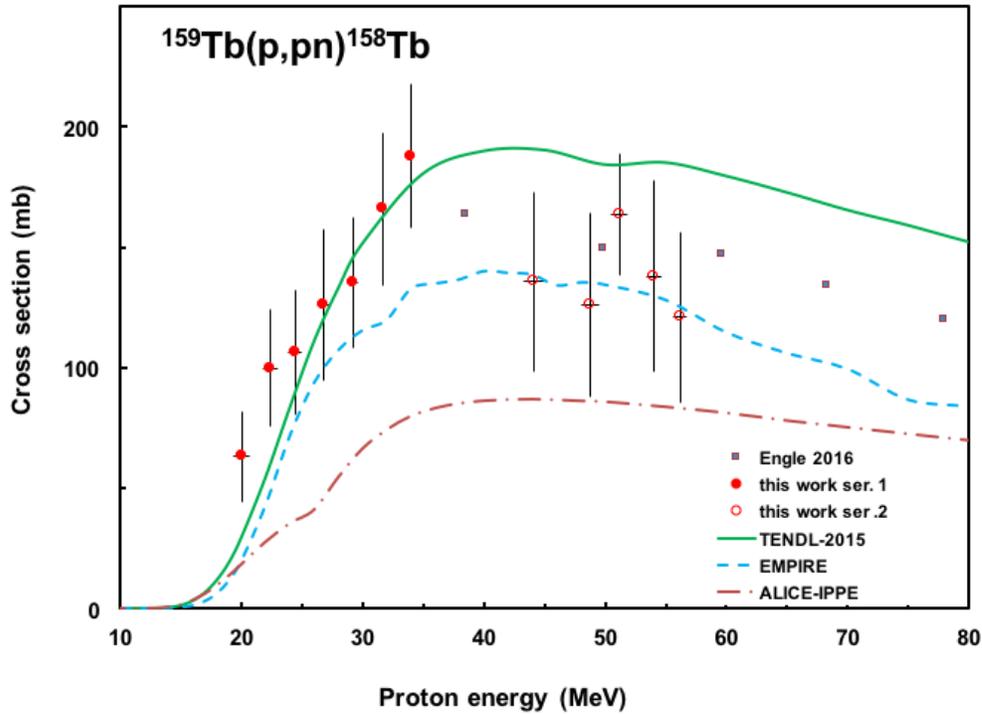

Fig. 5. Experimental and theoretical excitation functions of the $^{159}$Tb(p,pn)$^{158}$Tb reaction

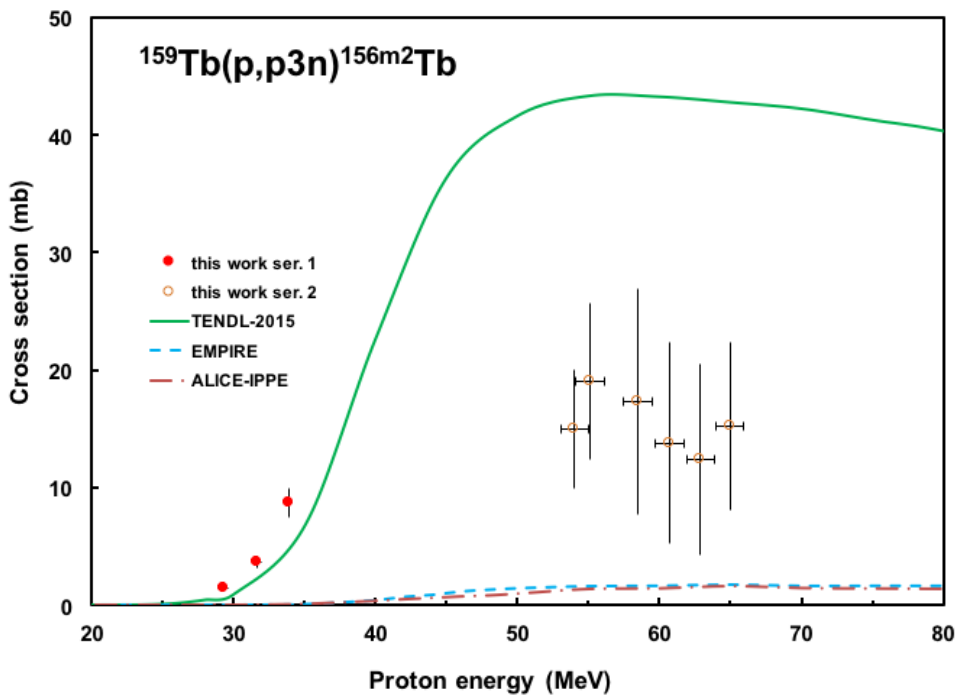

Fig. 6. Experimental and theoretical excitation functions of the $^{159}$Tb(p,p4n)$^{156m2}$Tb reaction

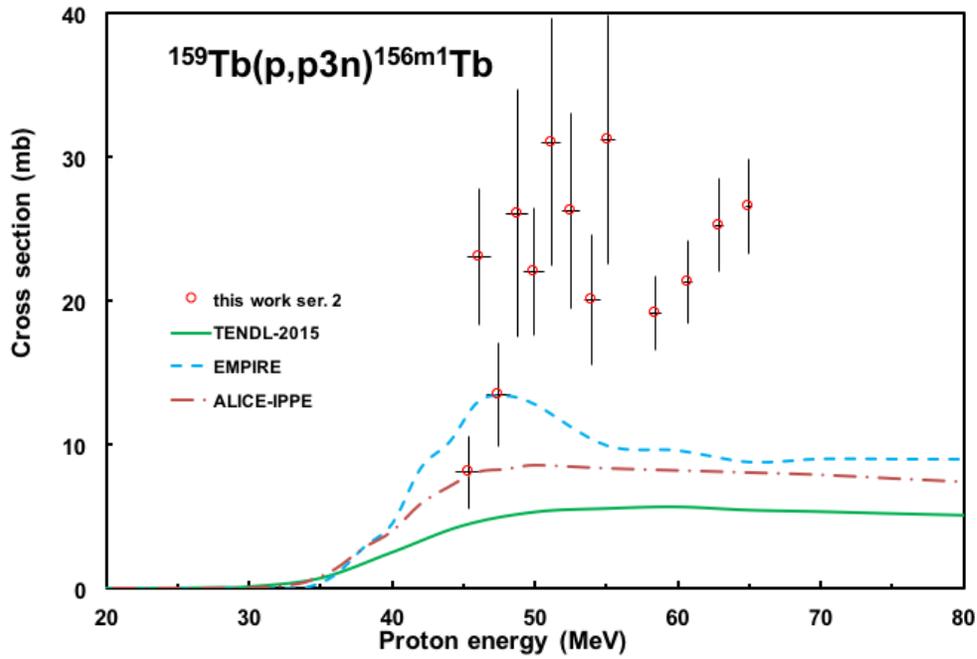

Fig. 7. Experimental and theoretical excitation functions of the $^{159}$Tb(p,p4n)$^{156m1}$Tb reaction

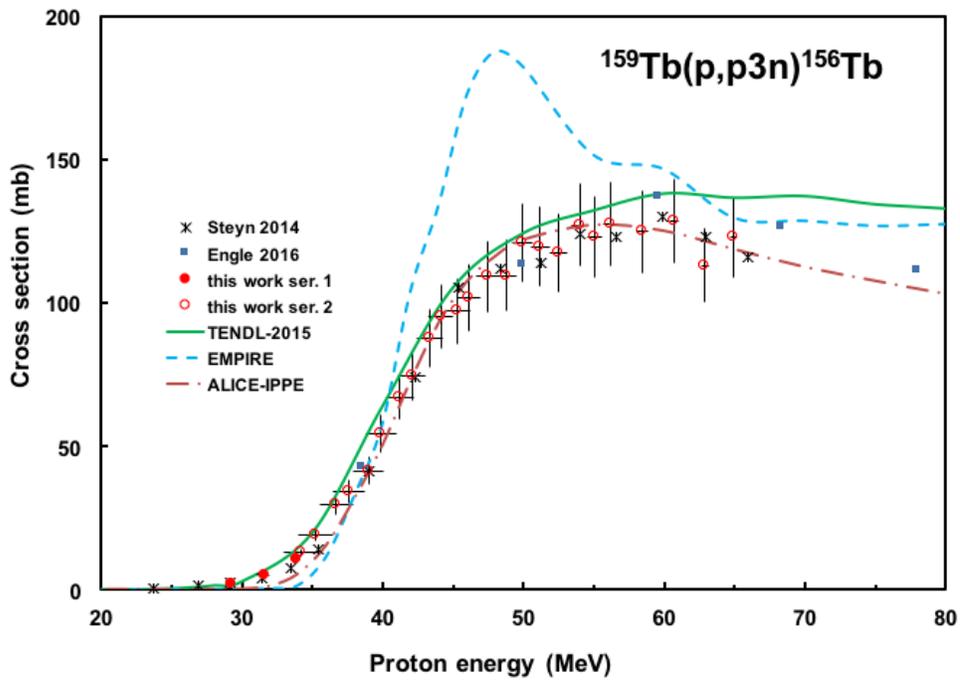

Fig. 8. Experimental and theoretical excitation functions of the $^{159}$Tb(d,p4n)$^{156g}$Tb(m+) reaction

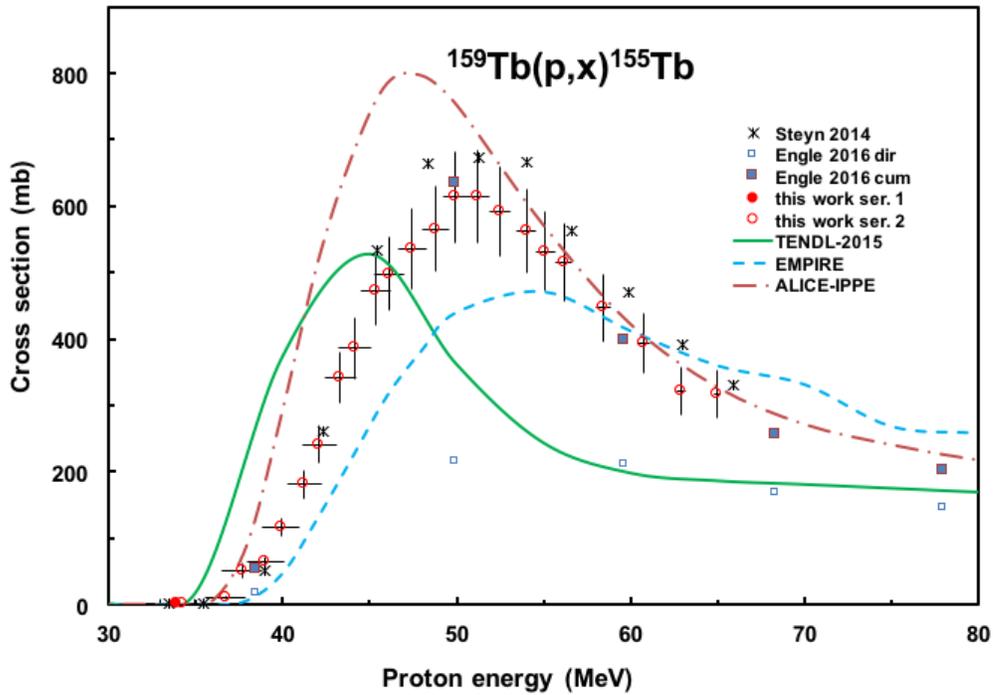

Fig. 9. Experimental and theoretical excitation functions of the $^{159}$Tb(p,x)$^{155}$Tb(cum) reaction

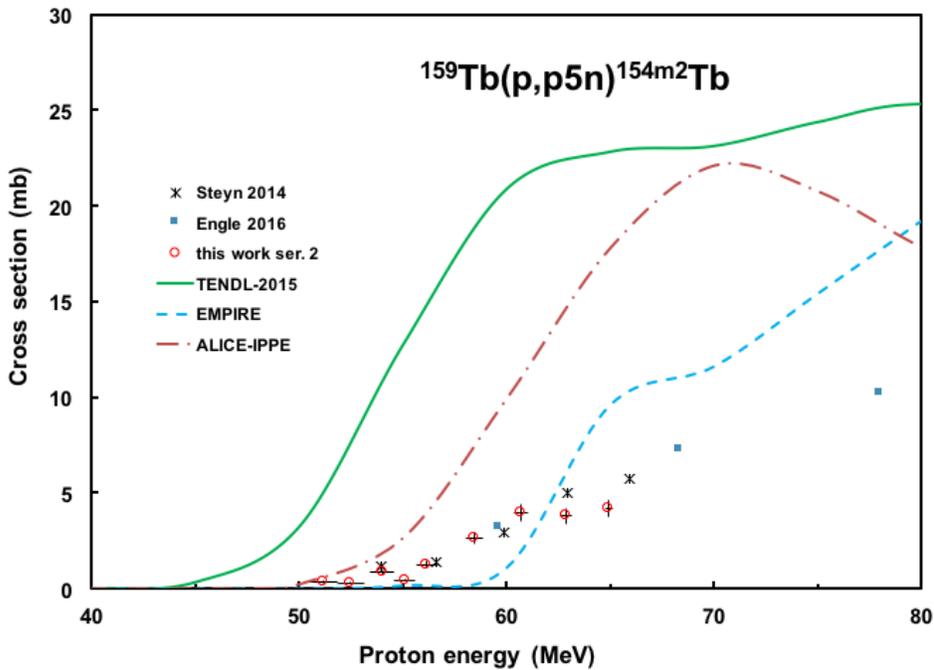

Fig. 10. Experimental and theoretical excitation functions of the $^{159}$Tb(p,p5n)$^{154m2}$Tb reaction

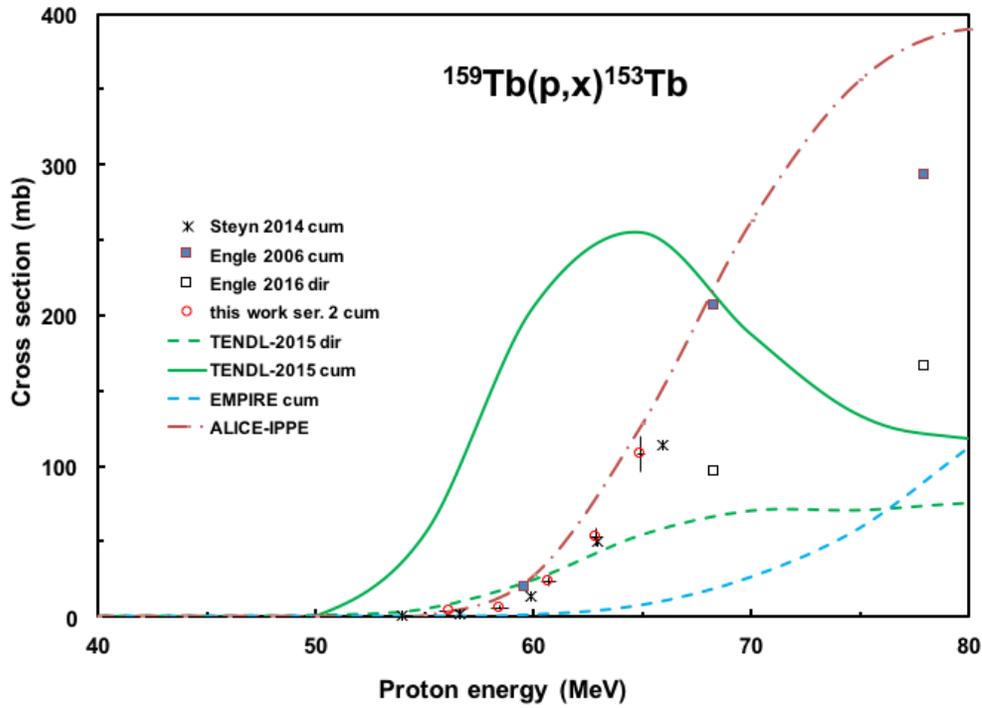

Fig. 11. Experimental and theoretical excitation functions of the $^{159}$Tb(p,x)$^{153}$Tb(cum) reaction

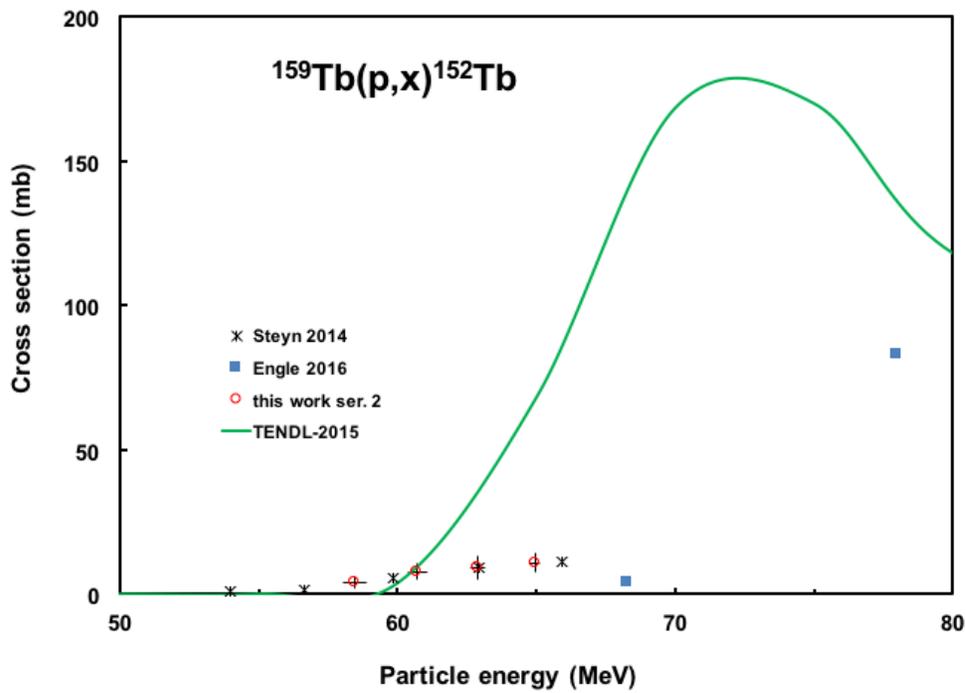

Fig. 12. Experimental and theoretical excitation functions of the $^{159}$Tb(p,x)$^{152}$Tb(cum) reaction

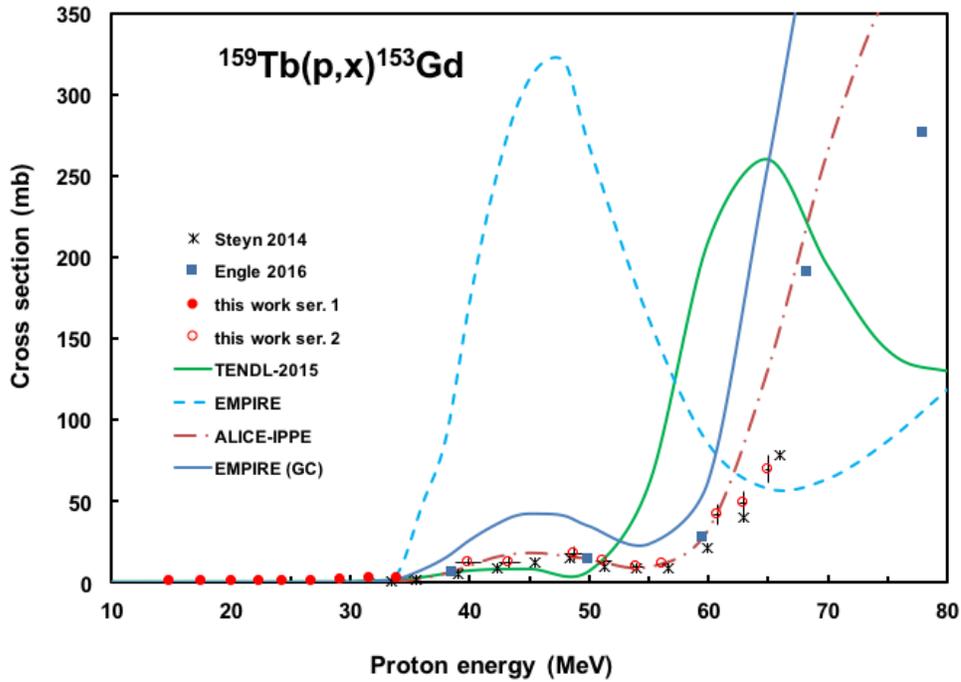

Fig. 13. Experimental and theoretical excitation functions of the $^{159}$Tb(d,x)$^{153}$Gd(cum) reaction

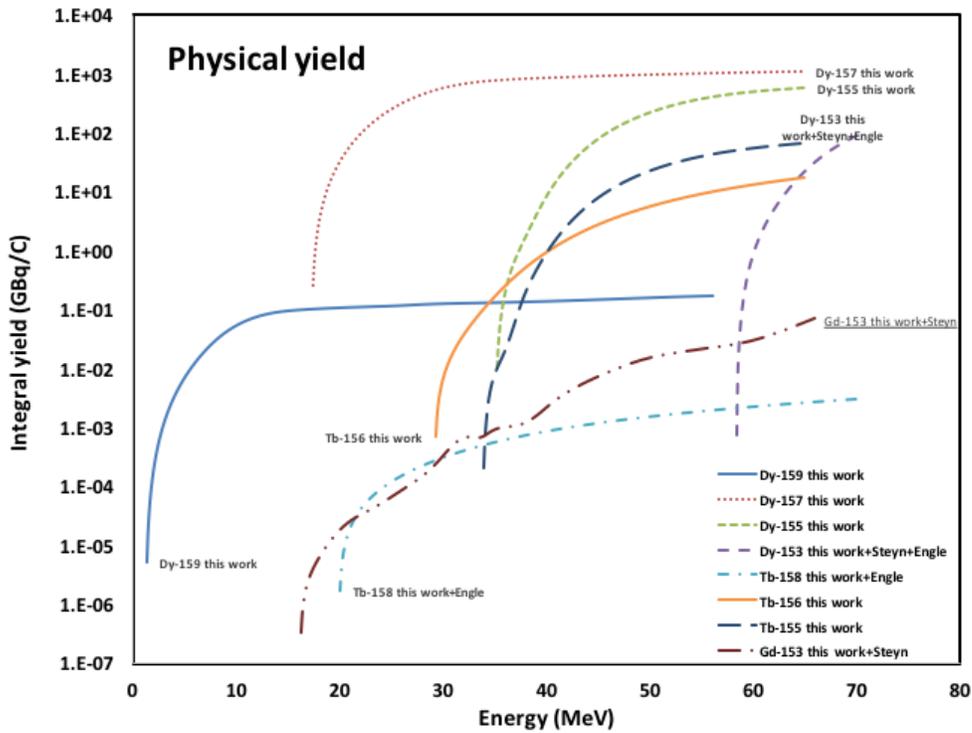

Fig. 14. Thick target yields for radionuclides of dysprosium, terbium and gadolinium